\documentclass[preprint,showpacs,superscriptaddress,nofootinbib]{revtex4}
\usepackage[dvips]{graphicx}
\usepackage{amsmath,amssymb}

\newcommand{\lromn}[1]{\uppercase\expandafter{\romannumeral#1}}

\newcommand{\MNS}{{\text{MNS}}}
\newcommand{\eV}{{\text{eV}}}
\newcommand{\MeV}{{\text{MeV}}}
\newcommand{\GeV}{{\text{GeV}}}
\newcommand{\TeV}{{\text{TeV}}}
\newcommand{\BR}{\text{BR}}

\newcommand{\BL}{{\text{B$-$L}}}
\newcommand{\DM}{{\text{DM}}}
\newcommand{\EM}{{\text{EM}}}

\newcommand{\U}{{\text{U}}}
\newcommand{\SU}{{\text{SU}}}
\newcommand{\Li}{{\text{Li}}}
\newcommand{\SI}{{\text{SI}}}

\newcommand{\SM}{{\text{SM}}}

\begin{document}

\preprint{UT-HET 089}
\preprint{MISC-2014-04}

\title{
 Neutrino Mass and Dark Matter
from Gauged $\U(1)_\BL$ Breaking
}

%%%%%%%%%%%%%%%%%%%%%%%%%%%%%%%%%%%%%%%%%%%%%%%%%%%%%%%%%%%%%%%%%%%%%%
\author{Shinya Kanemura}
\email{kanemu@sci.u-toyama.ac.jp}
\affiliation{
Department of Physics,
University of Toyama, Toyama 930-8555, Japan
}
%%%%%%%%%%%%%%%%%%%%%%%%%%%%%%%%%%%%%%%%%%%%%%%%%%%%%%%%%%%%%%%%%%%%%%
\author{Toshinori Matsui}
\email{matsui@jodo.sci.u-toyama.ac.jp}
\affiliation{
Department of Physics,
University of Toyama, Toyama 930-8555, Japan
}
%%%%%%%%%%%%%%%%%%%%%%%%%%%%%%%%%%%%%%%%%%%%%%%%%%%%%%%%%%%%%%%%%%%%%%
\author{Hiroaki Sugiyama}
\email{sugiyama@cc.kyoto-su.ac.jp}
\affiliation{
Maskawa Institute for Science and Culture,
Kyoto Sangyo University, Kyoto 603-8555, Japan
}
%%%%%%%%%%%%%%%%%%%%%%%%%%%%%%%%%%%%%%%%%%%%%%%%%%%%%%%%%%%%%%%%%%%%%%%

%%%%%% date %%%%%%%
%\date{\today}
%%%%%%%%%%%%%%%%%%%

\begin{abstract}
 We propose a new model
where the Dirac mass term for neutrinos,
the Majorana mass term for right-handed neutrinos,
and the other new fermion masses arise
via the spontaneous breakdown of the $\U(1)_\BL$ gauge symmetry.
 The anomaly-free condition gives
four sets of assignment of the $\BL$ charge to new particles,
and three of these sets have
an associated global $\U(1)_\DM$ symmetry
which stabilizes dark matter candidates.
 The dark matter candidates contribute
to generating the Dirac mass term for neutrinos
at the one-loop level.
 Consequently,
tiny neutrino masses are generated at the two-loop level
via a Type-I-Seesaw-like mechanism.
 We show that this model
can satisfy current bounds from
neutrino oscillation data,
the lepton flavor violation,
the relic abundance of the dark matter,
and the direct search for the dark matter.
 This model would be tested at future collider experiments
and dark matter experiments.
\end{abstract}

%14.60.Pq 	Neutrino mass and mixing
%12.60.Fr 	Extensions of electroweak Higgs sector
%14.70.Pw 	Other gauge bosons
%95.35.+d	Dark matter
%%%14.60.St 	Non-standard-model neutrinos, right-handed neutrinos, etc.

\maketitle

%%%%%%%%%%%%%%%%%%%%%%%%%%%%%
%%%%%  sec: intro  %%%%%%%%%%
%%%%%%%%%%%%%%%%%%%%%%%%%%%%%
\section{Introduction}
\label{sec:intro}

 The existence of neutrino masses
has been established very well
by the brilliant success of neutrino oscillation measurements%
~\cite{Ref:solar-v, Aharmim:2011vm, Ref:atm-v,
Ref:acc-disapp-v, Abe:2014ugx, Ref:acc-app-v,
Ref:short-reac-v, An:2013zwz, Ref:long-reac-v},
in spite that
neutrinos are massless in the standard model of particle physics~(SM)
where right-handed neutrinos $\nu_R^{}$ are absent.
 If $\nu_R^{}$ are introduced to the SM,
there are two possible mass terms for neutrinos~\cite{Ref:seesaw},
the Dirac type $\overline{\nu_L}\nu_R$
and the Majorana type $\overline{(\nu_R^{})^c} \nu_R^{}$.

 Since fermion masses in the SM
are generated via the spontaneous breakdown
of the $\SU(2)_L\times \U(1)_Y$ gauge symmetry,
it seems natural that
new fermion mass terms which do not exist in the SM
arise from spontaneous breakdown of a new gauge symmetry.
 Let us take a $\U(1)$ as the group of the new gauge symmetry
(denoted as $\U(1)^\prime$).
 Suppose that the $\U(1)^\prime$ gauge symmetry is
spontaneously broken by the vacuum expectation value~(VEV)
of a scalar field $\sigma^0$
which is a singlet under the SM gauge group.
 Then
origins of the Majorana mass term of $\nu_R^{}$
and the Dirac mass term of neutrinos
can be $\sigma^0 \overline{(\nu_R^{})^c} \nu_R^{}$%
~(or $(\sigma^0)^\ast \overline{(\nu_R^{})^c} \nu_R^{}$)
and $\sigma^0 \overline{\nu_R^{}} \Phi^T \epsilon L$,
respectively,
where the field $L$ is the $\SU(2)_L$-doublet of leptons,
$\Phi$ is the Higgs doublet field in the SM,
and $\epsilon$ is the complete antisymmetric tensor
for the $\SU(2)_L$ indices.
 The Majorana mass term for $\nu_L^{}$
comes from $(\sigma^0)^3 \overline{L^c} \epsilon \Phi^\ast \Phi^T \epsilon L$%
~(or $\sigma^0 |\sigma^0|^2 \overline{L^c} \epsilon \Phi^\ast \Phi^T \epsilon L$).

 When we decompose
the dimension-5 operator $\sigma^0 \overline{\nu_R^{}} \Phi^T \epsilon L$
with renormalizable interactions,
an interesting possibility is the radiative realization of the operator.
 A variety of models
where the Dirac mass term for neutrinos is radiatively generated
has been studied in Refs.~\cite{Mohapatra:1987hh, 1loopLR, 1loopE6,
Nasri:2001ax, Gu:2007ug, Wei:2010ww, Kanemura:2011jj, Okada:2014vla}
(See also Ref.~\cite{Babu:1989fg}).
 In a radiative mechanism for neutrino masses,
a dark matter candidate can appear
by imposing an {\it ad hoc} unbroken $Z_2$ symmetry%
~(See e.g., Refs~\cite{Gu:2007ug, Ref:KNT, Ref:Ma, Ma:2008cu, Ref:AKS,
Aoki:2011yk, Kanemura:2012rj, Ref:KNS, Kanemura:2013qva, Law:2013saa}).
 It would be natural that
such a symmetry to stabilize the dark matter
appears as a residual symmetry of a gauge symmetry
which is spontaneously broken at higher energies than
the electroweak scale%
~(See e.g., Refs.~\cite{Krauss:1988zc, Batell:2010bp, Ref:DiscreteGauge}).
 The breaking of such a gauge symmetry
can also be the origin of masses of new chiral fermions
which contribute to the loop diagram.
 If we take a $\U(1)^\prime$ symmetry as the new gauge symmetry
and introduced fermions are only singlet fields under the SM gauge group,
a simple choice for $\U(1)^\prime$ is the $U(1)_\BL$
because of the cancellation of
the $[\SU(3)_C]^2\times U(1)^\prime$,
the $[\SU(2)_L]^2\times U(1)^\prime$,
the $[\U(1)_Y]^2\times U(1)^\prime$
and the $\U(1)_Y \times [U(1)^\prime]^2$ anomalies.%
\footnote{
 See Ref.~\cite{Ma:2001kg}
for an anomaly-free $\U(1)^\prime$ gauge symmetry
when a new fermion field is introduced
as an $\SU(2)_L$-triplet with a hypercharge $Y=0$.
}
 New physics models with the TeV-scale $\U(1)_\BL$ gauge symmetry
can be found in e.g.,\ Refs.~\cite{Ref:BL, Ref:BL-2}.
 Collider phenomenology on the $\U(1)_\BL$ gauge symmetry
is discussed in e.g.,\ Ref.~\cite{Ref:BL-Pheno}.

 Along with the scenario stated above,
a model in Ref.~\cite{Ref:KNS} was constructed
such that the breaking of the $\U(1)_\BL$ gauge symmetry
gives a residual symmetry for the dark matter~(DM) stability
and new fermion mass terms which are absent in the SM
(e.g.,
the Majorana neutrino mass of $\nu_R^{}$,
the one-loop generated Dirac mass term of neutrinos,
and the masses of new fermions
among which the lightest one can be a DM candidate).
 However,
in order to cancel the anomalies
for the $\U(1)_\BL$ gauge symmetry,
it is required to introduce more new fermions
which do not contribute to the mechanism of generating neutrino masses.

 In this paper,
we propose a new model
which is an improved version of the model in Ref.~\cite{Ref:KNS}
from the view point of the anomaly cancellation.
 The $\BL$ charges of new particles
are assigned such that
the condition of anomaly cancellation is satisfied.
 Consequently,
the $\BL$ charges for some new particles
turn out to be irrational numbers.
 Because of this charge assignment,
there exists an unbroken global $\U(1)$ symmetry
even after the breakdown of the $\U(1)_\BL$ symmetry.
 The global $\U(1)$ symmetry
stabilizes the dark matter,
so that we hereafter call it $\U(1)_\DM$.
 The lightest particle with the irrational quantum number
can be a dark matter candidate.
 In our model,
the dark matter candidate
is a new scalar boson with the irrational quantum number.
 Furthermore,
the Dirac mass term of neutrinos
is radiatively generated
at the one-loop level
due to the quantum effect of the new particles
with irrational quantum numbers.
 Tiny neutrino masses
are explained by the two-loop diagrams
with a Type-I-Seesaw-like mechanism.
 We find that
the model can satisfy current data from
the neutrino oscillation, the lepton flavor violation~(LFV),
the relic abundance and the direct search for the dark matter,
and the LHC experiment.

 This paper is organized as follows.
 In Sec.~II,
the model is defined
and the basic property is discussed.
 In Sec.~III,
the neutrino masses are induced
due to the spontaneous breaking of $\U(1)_\BL$.
 We find a benchmark scenario
in which current experimental constraints
are taken into account
such as the neutrino oscillation data,
the LFV, the relic abundance of the dark matter,
the direct search for the dark matter,
and the LHC results.
 Conclusions are given in Sec.~IV\@.
 Some details of our calculations
are shown in Appendix.

%%%%%%%%%%%%%%%%%%%%%%%%%%%%%
%%%%%  sec: Model   %%%%%%%%%
%%%%%%%%%%%%%%%%%%%%%%%%%%%%%
\section{The model}
\label{sec:model}

%% tab: Particles >>>>>
\begin{table}[t]
\begin{center}
\begin{tabular}{c||c|c|c|c|c|c}
 {}
 & $s^0$
 & $\eta$
 & $\psi_{Ri}^{}$
 & $\psi_{Li}^{}$
 & $\nu_{Ra}^{}$
 & $\sigma^0$
\\\hline\hline
 Spin
 & $0$
 & $0$
 & \ $1/2$ \
 & $1/2$
 & $1/2$
 & $0$
\\\hline
 $\SU(2)_L$
 & {\bf \underline{1}}
 & {\bf \underline{2}}
 & {\bf \underline{1}}
 & {\bf \underline{1}}
 & {\bf \underline{1}}
 & {\bf \underline{1}}
\\\hline
 $\U(1)_Y$
 & $0$
 & $1/2$
 & $0$
 & $0$
 & $0$
 & $0$
\\\hline
 $\U(1)_\BL$
 & $x+1$
 & $x+1$
 & $x$
 & $x+2/3$
 & $-1/3$
 & \ $2/3$ \
\end{tabular}
\caption{
 Particle contents in this model.
 Indices $i$ (for $\psi_R^{}$ and $\psi_L^{}$) and $a$ (for $\nu_R^{}$)
run from $1$ to $N_\psi$ and from $1$ to $N_{\nu_R}$, respectively.
}
\label{tab:particle}
\end{center}
\end{table}
%%<<<<< tab: Particles

%% tab: B-L charge >>>>>
\begin{table}[t]
\begin{center}
\begin{tabular}{c||c|c|c|c}
 {}
 & Case I
 & Case II
 & Case III
 & Case IV
\\\hline\hline
 $N_\psi$
 & $1$
 & $2$
 & $3$
 & $4$
\\\hline
 $N_{\nu_R^{}}$
 & $7$
 & $5$
 & $3$
 & $1$
\\\hline
 $x$
 & $\frac{ 2\sqrt{3}-1 }{3}$
 & $\frac{ \sqrt{6}-1 }{3}$
 & $\frac{1}{\,3\,}$
 & $\frac{ \sqrt{3}-1 }{3}$
\end{tabular}
\caption{
 Sets of $N_\psi$, $N_{\nu_R^{}}$ and $x$,
for which the $\U(1)_\BL$ gauge symmetry is free from anomaly.
 Here,
$N_\psi$ is the number of $\psi_{Ri}^{}$
(the same as the number of $\psi_{Li}^{}$),
$N_{\nu_R^{}}$ is the number of $\nu_{R a}^{}$,
and $x$ is the $\BL$ charge of $\psi_{Ri}^{}$.
}
\label{tab:BL-charge}
\end{center}
\end{table}
%%<<<<< tab: B-L charge

 New particles listed in Table~\ref{tab:particle}
are added to the SM\@.
 Assignment of $\U(1)_\BL$ charges
is different from that in the previous model in Ref.~\cite{Ref:KNS}.
 Conditions for cancellation of
the $[\U(1)_\BL]\times[\text{gravity}]^2$ and $[\U(1)_\BL]^3$ anomalies are
%%----- eq: Anomaly Cancellation >>>>>
\begin{eqnarray}
3
-\frac{1}{\,3\,} N_{\nu_R^{}}
-\frac{2}{\,3\,} N_\psi
&=& 0 ,
\\
%%-----------------
3
-\frac{1}{27} N_{\nu_R^{}}
+\left( -2x^2 - \frac{4}{\,3\,} x - \frac{8}{27} \right) N_\psi
&=& 0 ,
\end{eqnarray}
%%<<<<< eq: Anomaly Cancellation -----
where $N_\psi$ is the number of $\psi_{Ri}^{}$
(the same as the number of $\psi_{Li}^{}$),
and $N_{\nu_R^{}}$ is the number of $\nu_{Ra}^{}$.
 There are four solutions as presented in Table~\ref{tab:BL-charge}.
 Except for Case~III,
the $\U(1)_\BL$ charges of some new particles are irrational numbers
while the $\U(1)_\BL$ symmetry is spontaneously broken
by the VEV of $\sigma^0$
whose $\U(1)_\BL$ charge is a rational number.
 Therefore,
the irrational charges are conserved,
and the lightest particle with an irrational $\U(1)_\BL$ charge
becomes stable
so that the particle can be regarded as a dark matter candidate.
 Notice that there is no dark matter candidate in Case~III\@.
 As we see later,
two of three light neutrinos are massless in Case~I\@,
which does not fit the neutrino oscillation data.
 In this paper,
we take Case~IV as an example.%
\footnote{
 If the $\BL$ charge of $\sigma^0$ is $2$
as in the model in Ref.~\cite{Ref:KNS},
the $\BL$ charges for
$\{ s^0 ,\, \eta ,\, \psi_R^{} ,\, \psi_L^{} ,\, \nu_R^{} \}$
will be assigned as
$\{ x+1 ,\, x+1 ,\, x ,\, x+2 ,\, -1 \}$.
 There is only an anomaly-free solution $x=-1$.
 We do not take this possibility
because there is no residual symmetry
to stabilize the dark matter.
}

 The Yukawa interactions are given by
%%----- eq: Yukawa >>>>>
\begin{eqnarray}
 {\cal L}_{\text{Yukawa}}
 &=&
 {\cal L}_{\text{SM-Yukawa}}
 -
 (y_R^{})_a\,
 \overline{ (\nu_R^{})_a }\, (\nu_R^{})^c_a\, (\sigma^0)^\ast
 -
 (y_{\Psi}^{})_i\,
 \overline{ (\psi_R)_i }\, (\psi_L)_i\, (\sigma^0)^\ast
 \nonumber\\
 &&\hspace*{-2mm}
{}-
 h_{ia}\,
 \overline{ (\psi_L)_i }\, (\nu_R^{})_a\, s^0
 -
 f_{\ell i}\,
 \overline{ L_\ell }\, (\psi_R)_i\, \tilde{\eta}
 + \text{h.c.} ,
\label{eq:Yukawa}
\end{eqnarray}
%%<<<<< eq: Yukawa -----
where ${\cal L}_{\text{SM-Yukawa}}$ denotes
the Yukawa interactions in the SM,
$L_\ell$~($\ell = e, \mu, \tau)$
are the $\SU(2)_L$ doublet fields of the SM leptons,
and $\tilde{\eta} \equiv ( (\eta^0)^\ast, -\eta^- )^T$.
 Indices $i$ and $a$ run
from $1$ to $N_\psi$ and from $1$ to $N_{\nu_R^{}}$,
respectively.
 Notice that
a Yukawa interaction $\overline{(\nu_R)_a^c} \psi_{Ri}^{} (s^0)^\ast$
which exists in the previous model is absent in this model
because of assignment of $\BL$ charge to new particles
are different from those in the previous paper~\cite{Ref:KNS}.

 The scalar potential in our model
is the same as that in the previous model%
\footnote{
 For Case~III in Table~\ref{tab:BL-charge},
there are additional terms
e.g.\ $(s^0)^\ast (\sigma^0)^2$.
 See also Ref.~\cite{Basso:2012ti}.
}~\cite{Ref:KNS}:
%%----- eq: Scalar potential >>>>>
\begin{eqnarray}
 V(\Phi, s, \eta, \sigma)
 &=&
 -\mu_{\phi}^2\Phi^{\dagger} \Phi
 +
 \mu_s^2 |s^0|^2
 +
 \mu_{\eta}^2\eta^{\dagger} \eta
 -
 \mu_{\sigma}^2 |\sigma^0|^2
 \nonumber\\
 &&{}
 +
 \lambda_\phi \left(\Phi^{\dagger} \Phi\right)^2
 +
 \lambda_s |s^0|^4
 +
 \lambda_\eta \left(\eta^{\dagger} \eta\right)^2
 +
 \lambda_\sigma |\sigma^0|^4
 \nonumber\\
 &&{}
 +
 \lambda_{s\eta} |s^0|^2 \eta^{\dagger} \eta
 +
 \lambda_{s\phi} |s^0|^2 \Phi^{\dagger} \Phi
 +
 \lambda_{\phi\phi} (\eta^{\dagger}\eta) (\Phi^{\dagger} \Phi)
 +
 \lambda_{\eta\phi} (\eta^{\dagger} \Phi) (\Phi^{\dagger}\eta)
 \nonumber\\
 &&{}
 +
 \lambda_{s\sigma} |s^0|^2 |\sigma^0|^2
 +
 \lambda_{\sigma\eta} |\sigma^0|^2 \eta^{\dagger} \eta
 +
 \lambda_{\sigma\phi} |\sigma^0|^2 \Phi^{\dagger} \Phi
 +\left( \mu_3^{}\, s^0\, \eta^{\dagger}\, \Phi + \text{h.c.}\right) ,
 \label{eq:V}
\end{eqnarray}
%%<<<<< eq: Scalar potential -----
where $\mu_{\phi}^2$, $\mu_s^2$, $\mu_{\eta}^2$, and $\mu_{\sigma}^2$
are defined as positive values.
Without loss of generality,
we can take a real positive $\mu_3^{}$
by utilizing a rephasing of $s^0$.

 Two scalar fields $\phi^0$ and $\sigma^0$ obtain VEVs
$v_\phi^{}$~[$= \sqrt{2}\, \langle \phi^0 \rangle = 246\,\GeV$]
and $v_\sigma^{}$~[$= \sqrt{2}\, \langle \sigma^0 \rangle$].
 Then
$\SU(2)_L\times\U(1)_Y$ and $\U(1)_\BL$ gauge symmetries
are spontaneously broken by $v_\phi^{}$ and $v_\sigma^{}$,
respectively.
 These VEVs are given by
%%----- eq: vev >>>>>
\begin{eqnarray}
\begin{pmatrix}
 v_\phi^2\\
 v_\sigma^2  
\end{pmatrix}
=
 \frac{1}{\lambda_\sigma \lambda_\phi - \lambda_{\sigma\phi}^2/4}
 \begin{pmatrix}
  \lambda_\sigma          & -\lambda_{\sigma\phi}/2\\
  -\lambda_{\sigma\phi}/2 & \lambda_\phi
 \end{pmatrix}
 \begin{pmatrix}
  \mu_\phi^2\\
  \mu_\sigma^2
 \end{pmatrix} .
\end{eqnarray}
%%<<<<< eq: vev -----
 The VEV $v_\sigma^{}$ provides
a mass of the $\U(1)_\BL$ gauge boson $Z^\prime$
as $m_{Z^\prime}^{} = (2/3) g_\BL^{} v_\sigma^{}$,
where $g_\BL^{}$ is the $\U(1)_\BL$ gauge coupling constant.
 After the gauge symmetry breaking
with $v_\phi^{}$ and $v_\sigma^{}$,
we can confirm in Eqs.~\eqref{eq:Yukawa} and \eqref{eq:V}
that there is a residual global $\U(1)_\DM$ symmetry,
for which irrational $\U(1)_\BL$-charged particles
($\eta$, $s^0$, $\psi_{Li}^{}$, and $\psi_{Ri}^{}$)
have the same $\U(1)_\DM$-charge while the other particles are neutral.

 We have two CP-even scalar particles
$h^0$ and $H^0$ as
%%----- eq: theta0 >>>>>
\begin{eqnarray}
\begin{pmatrix}
 h^0\\
 H^0
\end{pmatrix}
=
 \begin{pmatrix}
  \cos\theta_0 & -\sin\theta_0\\
  \sin\theta_0 & \cos\theta_0
 \end{pmatrix}
 \begin{pmatrix}
  \phi^0_r\\
  \sigma^0_r
 \end{pmatrix} , \quad
%
%----------------
%
\sin{2\theta_0}
=
 \frac{ 2 \lambda_{\sigma\phi}^{}\, v_\phi\, v_\sigma^{} }
      { m_{H^0}^2 - m_{h^0}^2 } ,
\end{eqnarray}
%%<<<<< eq: theta0 -----
where
$\phi^0 = ( v_\phi^{} + \phi^0_r + i z_\phi^{} )/\sqrt{2}$
and
$\sigma^0 = ( v_\sigma^{} + \sigma^0_r + i z_\sigma^{} )/\sqrt{2}$.
 Nambu-Goldstone bosons $z_\phi^{}$ and $z_\sigma^{}$
are absorbed by $Z$ and $Z^\prime$ bosons, respectively.
 Masses of $h^0$ and $H^0$ are given by
%%----- eq: m0 >>>>>
\begin{eqnarray}
m_{h^0}^2
&=&
 \lambda_\phi v_\phi^2 + \lambda_\sigma v_\sigma^2
 -\sqrt{
   \left(
    \lambda_\phi v_\phi^2 - \lambda_\sigma v_\sigma^2
   \right)^2
   + \lambda_{\sigma\phi}^2 v_\phi^2 v_\sigma^2 }\, ,
 \nonumber\\
%
%-----------------------
%
m_{H^0}^2
&=&
 \lambda_\phi v_\phi^2 + \lambda_\sigma v_\sigma^2
 +\sqrt{
   \left(
    \lambda_\phi v_\phi^2 - \lambda_\sigma v_\sigma^2
   \right)^2
   + \lambda_{\sigma\phi}^2 v_\phi^2 v_\sigma^2 }\, .
\end{eqnarray}
%%<<<<< eq: m0 -----
 On the other hand,
$\eta^0$ and $s^0$ do not mix with
$\phi^0$ and $\sigma^0$
even though the $\U(1)_\BL$ symmetry is broken by $v_\sigma^{}$.
 Two neutral complex scalars ${\mathcal H}^0_1$ and ${\mathcal H}^0_2$
are obtained by
%%----- eq: theta0' >>>>>
\begin{eqnarray}
\begin{pmatrix}
 {\mathcal H}_1^0\\
 {\mathcal H}_2^0
\end{pmatrix}
=
 \begin{pmatrix}
  \cos\theta_0^\prime & -\sin\theta_0^\prime\\
  \sin\theta_0^\prime & \cos\theta_0^\prime
 \end{pmatrix}
 \begin{pmatrix}
  \eta^0\\
  s^0
 \end{pmatrix} , \quad
%
%----------------"
%
\sin{2\theta_0^\prime}
=
 \frac{ \sqrt{2}\, \mu_3^{}\, v_\phi^{} }
      { m_{{\mathcal H}_2^0}^2 - m_{{\mathcal H}_1^0}^2 } .
\end{eqnarray}
%%<<<<< eq: theta0' -----
 Their masses and the mass of the charged scalar $\eta^\pm$
are given by
%%----- eq: mDM >>>>
\begin{eqnarray}
m_{{\mathcal H}_1^0}^2
&=&
 \frac{1}{2}
 \left(
  m_\eta^2 + m_s^2
  -\sqrt{ \left( m_\eta^2 - m_s^2 \right)^2 + 2 \mu_3^2 v_\phi^2 }
 \right) ,
\\
%
%--------------------------
%
m_{{\mathcal H}_2^0}^2
&=&
 \frac{1}{2}
 \left(
  m_\eta^2 + m_s^2
  +\sqrt{ \left( m_\eta^2 - m_s^2 \right)^2 + 2 \mu_3^2 v_\phi^2 }
 \right) ,
\\
%
%--------------------------
%
m_{\eta^{\pm}}^2
&=&
 m_\eta^2 - \lambda_{\eta\phi} \frac{v_\phi^2}{\,2\,} ,
\end{eqnarray}
%%<<<<< eq: mDM -----
where
$m_s^2
\equiv
 \mu_s^2
 + \lambda_{s\phi} v_\phi^2/2
 + \lambda_{s\sigma} v_\sigma^2/2$
and
$m_{\eta}^2
\equiv
 \mu_{\eta}^2
 + \left( \lambda_{\phi\phi}
 + \lambda_{\eta\phi} \right) v_\phi^2/2
 + \lambda_{\sigma\eta} v_\sigma^2/2$.

%%%%%%%%%%%%%%%%%%%%%%%%%%%%%
%%%  sec: numass and DM   %%%
%%%%%%%%%%%%%%%%%%%%%%%%%%%%%
\section{Neutrino Mass and Dark Matter}
\label{sec:nuDM}

%%%%%%%%%%%%%%%%%%%%%%%%
%%%  subsec: numass  %%%
%%%%%%%%%%%%%%%%%%%%%%%%
\subsection{Neutrino Mass}
\label{subsec:numass}

%%----- fig: Two-loop diagrams >>>>>
\begin{figure}[t]
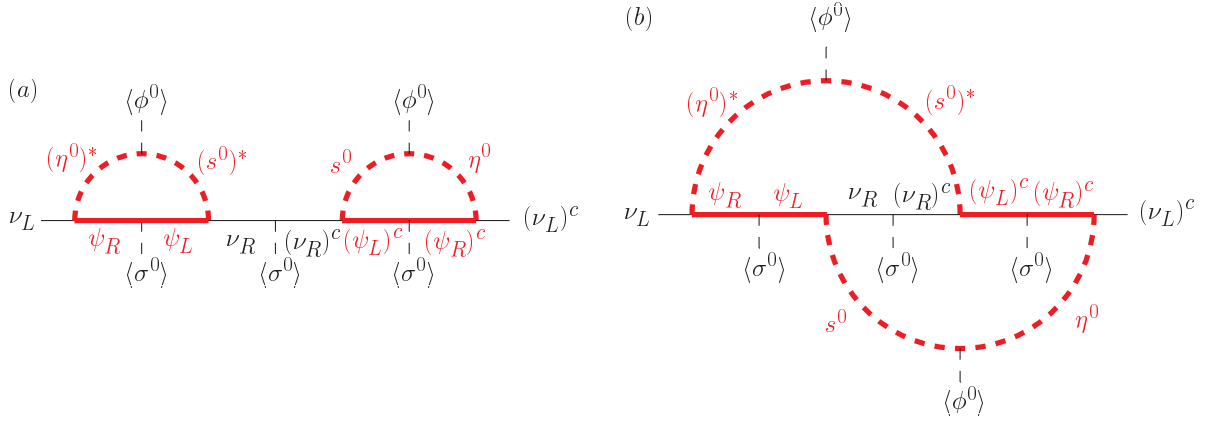

\begin{minipage}[b]{80mm}
\includegraphics[width=7.7cm]{numass-1.eps}
\vspace*{17mm}
\end{minipage}
\begin{minipage}[t]{80mm}
\includegraphics[width=7.7cm]{numass-2.eps}
\end{minipage}
\caption{
 Two-loop diagrams for tiny neutrino masses in this model.
 Bold (red) lines are propagators
of particles of irrational $\U(1)_\BL$ charges.
}
\label{Fig:two-loop}
\end{figure}
%%<<<<< fig: Two-loop diagrams -----

 Tiny neutrino masses are
generated by two-loop diagrams
in Fig.~\ref{Fig:two-loop}~\cite{Ref:KNS}.
 The mass matrix $m_\nu^{}$
is expressed in the flavor basis as
%%----- eq: Neutrino mass matrix >>>>>
\begin{eqnarray}
(m_\nu)_{\ell\ell^\prime}^{}
=
 \frac{1}{ \left(16\pi^2\right)^2 }
 \sum_{i,j,a}
 f_{\ell i}\, h_{ia}\, (m_R^{})_a\, (h^T)_{aj}\, (f^T)_{j\ell^\prime}
 \Bigl\{
  (I_1)_{ija} + (I_2)_{ija}
 \Bigr\} .
\label{eq:mnu}
\end{eqnarray}
%%<<<<< eq: Neutrino mass matrix -----
 Loop functions
$(I_1)_{ija}$ and $(I_2)_{ija}$
correspond to contributions of diagram (a) and (b)
in Fig.~\ref{Fig:two-loop},
respectively.
 Explicit forms of these loop functions
are shown in Appendix~\ref{sec:loop}.

 Let us define the following matrix:
%%----- eq: loop matrix >>>>>
\begin{eqnarray}
A_{ij}
&\equiv&
 \sum_a
 h_{ia} (m_R^{})_a (h^T)_{aj}
 \Bigl\{
  (I_1)_{ija} + (I_2)_{ija}
 \Bigr\} .
\label{eq:loop-matrix}
\end{eqnarray}
%%<<<<< eq: loop matrix -----
 If $N_\psi = 1$,
the matrix $A_{ij}$ becomes just a number
and then $(m_\nu)_{\ell\ell^\prime}^{}$ becomes
a rank-1 matrix which is not consistent with
neutrino oscillation data.
 Therefore,
Case~I in Table~\ref{tab:BL-charge}
is not acceptable.
 On the other hand,
$N_{\nu_R^{}} = 1$ does not mean
that $(m_\nu)_{\ell\ell^\prime}^{}$ is a rank-1 matrix
because of existence of $(I_2)_{ija}$.
 We will see later
that our benchmark point for Case~IV in Table~\ref{tab:BL-charge}
does not include massless neutrinos
even though $N_{\nu_R^{}} = 1$.

 The neutrino mass matrix $(m_\nu)_{\ell\ell^\prime}^{}$
is diagonalized by a unitary matrix $U_\MNS$,
the so-called Maki-Nakagawa-Sakata~(MNS) matrix~\cite{Maki:1962mu}, as
$U_\MNS^\dagger\, m_\nu\, U_\MNS^\ast
=
 \text{diag}( m_1 e^{i\alpha_1} ,\,
 m_2 e^{i\alpha_2} ,\, m_3 e^{i\alpha_3} )$.
 We take $m_i$~($i= 1\text{-}3$) to be real and positive values.
 Two differences of three phases $\alpha_i$
are physical Majorana phases~\cite{Ref:M-Phase}.
 The MNS matrix can be parametrized as
%%----- eq: MNS >>>>>
\begin{eqnarray}
U_\MNS
=
 \begin{pmatrix}
  1 & 0 & 0\\
  0 & c_{23} & s_{23}\\
  0 & -s_{23} & c_{23}
 \end{pmatrix}
 \begin{pmatrix}
  c_{13} & 0 & s_{13} e^{-i\delta}\\
  0 & 1 & 0\\
  -s_{13} e^{i\delta} & 0 & c_{13}
 \end{pmatrix}
 \begin{pmatrix}
  c_{12} & s_{12} & 0\\
  -s_{12} & c_{12} & 0\\
  0 & 0 & 1
 \end{pmatrix} ,
\end{eqnarray}
%%<<<<< eq: MNS -----
where $c_{ij} \equiv \cos\theta_{ij}$
and $s_{ij} \equiv \sin\theta_{ij}$.
 In our analysis,
the following values~\cite{Abe:2014ugx,An:2013zwz,Aharmim:2011vm}
obtained by neutrino oscillation measurements
are used in order to search for a benchmark point
of model parameters:
%%----- eq: nu params >>>>>
\begin{eqnarray}
m_1
&=&
 10^{-4}\,\eV ,
\\
%
%------------
%
\Delta m^2_{21}
&=&
 7.46\times 10^{-5}\,\eV^2 ,\,
%
%------------
%
\Delta m^2_{32}
=
 +2.51\times 10^{-3}\,\eV^2 ,
\\
%
%------------
%
\sin^2{2\theta_{23}}
&=&
 1 ,\
%
%------------
%
\sin^2{2\theta_{13}}
=
 0.09 ,\
%
%------------
%
\tan^2\theta_{12}
=
 0.427 ,\\
%
%------------
%
\delta
&=&
 0 ,\
%
%------------
%
\bigl\{
 \alpha_1 ,\, \alpha_2 ,\, \alpha_3
\bigr\}
=
 \bigl\{
  0 ,\, 0 ,\, 0
 \bigr\} ,
\end{eqnarray}
%%<<<<< eq: nu params -----
where $\Delta m^2_{ij} \equiv m_i^2 - m_j^2$.
 By using an ansatz presented in Appendix~\ref{sec:ansatz}
for the structure of Yukawa matrix $f_{\ell i}$,
we found a benchmark point as
%%----- eq: benchmark >>>>>
\begin{eqnarray}
&&
f
=
 \begin{pmatrix}
  1.79
   & -2.49
   & -1.97
   & 2.56\\
  -1.82
   & 1.10
   & 1.30
   & -0.818\\
  1.40
   & -0.598
   & -0.905
   & 0.222
 \end{pmatrix} \times 10^{-2},
%f
%=
% \begin{pmatrix}
%  1.78686
%   & -2.48746
%   & -1.9737
%   & 2.55808\\
%%
%  -1.82223
%   & 1.10461
%   & 1.29624
%   & -0.818099\\
%%
%  1.40402
%   & -0.598335
%   & -0.904845
%   & 0.222417
% \end{pmatrix} \times 10^{-2},
\label{eq:f-benchmark}
\\
%
%------------
%
&&
h
=
 \begin{pmatrix}
  0.7
   & 0.8
   & 0.9
   & 1
 \end{pmatrix}^T,
\\
%
%------------
%
&&
(m_R)_1
=
 250\,\GeV,
\\
%
%------------
%
&&
\bigl\{
 m_{\psi_1}^{} ,\, m_{\psi_2}^{} ,\,
 m_{\psi_3}^{} ,\, m_{\psi_4}^{}
\bigr\}
=
 \bigl\{
  650\,\GeV ,\, 750\,\GeV ,\, 850\,\GeV ,\, 950\,\GeV
 \bigr\} ,
\\
%
%------------
%
&&
\bigl\{
 m_{h^0}^{} ,\, m_{H^0}^{} ,\, \cos\theta_0
\bigr\}
=
 \bigl\{ 125\,\GeV ,\, 1000\,\GeV ,\, 1 \bigr\} ,
\label{eq:neutral-benchmark}
\\
%
%------------
%
&&
\bigl\{
 m_{{\mathcal H}_1^0}^{} ,\, m_{{\mathcal H}_2^0}^{} ,\, \cos\theta_0^\prime
\bigr\}
=
 \bigl\{ 60\,\GeV ,\, 450\,\GeV ,\, 0.05 \bigr\} ,
\\
%
%------------
%
&&
m_{\eta^\pm}^{}
=
 420\,\GeV ,
\\
%
%------------
%
&&
\bigl\{
 g_\BL^{} ,\, m_{Z^\prime}^{}
\bigr\}
=
 \bigl\{ 0.1 ,\, 4000\,\GeV \bigr\} .
\label{eq:BL-benchmark}
\end{eqnarray}
%%<<<<< tab: benchmark -----
 The values of $\{ g_\BL^{} ,\, m_{Z^\prime}^{}\}$
mean $v_\sigma^{} = 60\,\TeV$.
 The values of $\{ m_{h^0}^{} ,\, m_{H^0}^{} ,\, \cos\theta_0 \}$
correspond to
$\lambda_\phi \simeq 0.13$,
$\lambda_\sigma \simeq 2.8\times 10^{-4}$
and $\lambda_{\sigma\phi} = 0$.
 The values of
$\{ m_{{\mathcal H}_1^0}^{} ,\, m_{{\mathcal H}_2^0}^{} ,\,
 \cos\theta_0^\prime \}$
and $m_{\eta^\pm}^{}$
can be produced by
$m_s^{} \simeq 60\,\GeV$,
$m_\eta^{} \simeq 450\,\GeV$,
$\mu_3^{} \simeq 57\,\GeV$
and $\lambda_{\eta\phi} \simeq 0.86$.

%%%%%%%%%%%%%%%%%%%%%%%
%%%  subsec: LFV   %%%%
%%%%%%%%%%%%%%%%%%%%%%%
\subsection{Lepton Flavor Violation}
\label{subsec:LFV}

 The charged scalar $\eta^\pm$
contributes to the LFV decays
of charged leptons.
 The formula for the branching ratio~(BR) of $\mu \to e\gamma$
can be calculated~\cite{Hisano:1995cp} as
%%<<<<< eq: BR(meg) -----
\begin{eqnarray}
\BR(\mu\to e \gamma)
&=&
 \frac{3\alpha_\EM^{}}{64\pi G_F^2}
 \left|
  \frac{1}{m_{\eta^{\pm}}^2} f_{\mu i}\,
  F\!\left(
      \frac{ m_{\psi_i}^2 }{ m_{\eta^{\pm}}^2 }
     \right) (f^\dagger)_{ie}
 \right|^2 ,
 \label{eq:meg}
\end{eqnarray}
%%<<<<< eq: BR(meg) -----
where
%%<<<<< eq: F_meg -----
\begin{eqnarray}
F\!\left(x\right)
&\equiv&
 \frac{ 1 - 6x + 3x^2 + 2x^3 - 6x^2 \ln(x) }
      { 6\left( 1-x \right)^4 }.
\end{eqnarray}
%%<<<<< eq: F_meg -----
 At the benchmark point,
we have $\BR(\mu\to e\gamma) = 6.1\times 10^{-14}$
%$\BR(\mu\to e\gamma) = 6.06545\times 10^{-14}$
which satisfies the current constraint
$\BR(\mu\to e\gamma) < 5.7\times 10^{-13}$~(90\%~C.L.)~\cite{Ref:MEG}.

%%%%%%%%%%%%%%%%%%%%%%%
%%%  subsec: DM   %%%%
%%%%%%%%%%%%%%%%%%%%%%%
\subsection{Dark Matter}
\label{subsec:DM}

 In principle,
$\psi_1$ or ${\mathcal H}^0_1$
can be a dark matter candidate.
 However,
due to the following reason,
the scalar ${\mathcal H}^0_1$
turns out to be the dark matter candidate.
 If the dark matter is the fermion $\psi_1$,
it annihilates into a pair of SM particles
via the $s$-channel process mediated by $h^0$ and $H^0$.
 The cross section of the process
is proportional to $\sin^2{2\theta_0}$.
 In order to obtain a sufficient annihilation cross section of $\psi_1$,
a large mixing $\cos\theta_0 \simeq 1/\sqrt{2}$ is preferred%
~\cite{Ref:BL-2}.
 Even for a maximal mixing $\cos\theta_0 = 1/\sqrt{2}$,
the observed abundance of the dark matter~\cite{Ade:2013zuv}
requires $v_\sigma^{} \lesssim 10\,\TeV$.
 The current constraint from direct searches of the dark matter~\cite{Ref:LUX}
requires larger $v_\sigma^{}$
in order to suppress the $Z^\prime$ contribution.%
\footnote{
 This is because $m_{Z^\prime}^{}/g_\BL$
is not $2 v_\sigma$ as usual but $2 v_\sigma/3$ in this model.
}

 Because of the tiny mixing $\cos\theta_0^\prime = 0.05$,
the scalar dark matter ${\mathcal H}^0_1$ at the benchmark point
is dominantly made from $s^0$
which is a gauge-singlet field under the SM gauge group.
 The  annihilation of ${\mathcal H}^0_1$ into a pair of the SM particles
is dominantly caused by the $s$-channel scalar mediation via $h^0$%
~\cite{Kanemura:2010sh}
because $H^0$ is assumed to be heavy.
 The coupling constant $\lambda_{{\mathcal H}^0_1 {\mathcal H}^0_1 h^0}^{}$ for
the $\lambda_{{\mathcal H}^0_1 {\mathcal H}^0_1 h^0}^{}\,
v_\phi^{} {\mathcal H}^0_1 {\mathcal H}^{0\ast}_1 h^0$ interaction
controls the annihilation cross section,
the invisible decay $h^0 \to {\mathcal H}^0_1 {\mathcal H}^{0\ast}_1$
in the case of kinematically accessible,
and the $h^0$ contribution to
the spin-independent scattering cross section $\sigma_\SI$ on a nucleon.
 In Ref.~\cite{Cline:2013gha}, for example,
we see that ${\mathcal H}^0_1$
with $m_{{\mathcal H}^0_1}^{} = 60\,\GeV$ and
$\lambda_{{\mathcal H}^0_1 {\mathcal H}^0_1 h^0}^{} \sim 10^{-3}$
can satisfy constraints
from the relic abundance of the dark matter
and the invisible decay of $h^0$.
 We see also that the $h^0$ contribution to $\sigma_\SI$
is small enough to satisfy the current constraint
$\sigma_\SI < 9.2\times 10^{-46}\,\text{cm}^2$
for $m_\DM^{} = 60\,\GeV$~\cite{Ref:LUX}.
 Although the scattering of ${\mathcal H}^0_1$ on a nucleon
is mediated also by the $Z^\prime$ boson in this model,
the contribution can be suppressed
by taking a large $v_\sigma^{}$.
 The benchmark point corresponds to $v_\sigma^{} = 60\,\TeV$
and gives about $6.6\times 10^{-47}\,\text{cm}^2$
%$6.59289\times 10^{-47}\,\text{cm}^2$
for the scattering cross section via $Z^\prime$,
which is smaller than the current constraint~\cite{Ref:LUX}
by an order of magnitude.
 Thus,
the constraint from the direct search of the dark matter
is also satisfied at the benchmark point.

%%%%%%%%%%%%%%%%%%%%%%%%%%%%
%%%  subsec: collider   %%%%
%%%%%%%%%%%%%%%%%%%%%%%%%%%%
\subsection{Collider Phenomenology}
\label{subsec:coll}

%%---- tab: Z' decay >>>>>
\begin{table}[t]
\begin{center}
\begin{tabular}{c|c|c|c|c|c|c|c|c|c|c}
 \ \ \ $q\, \overline{q}$ \ \ \
 & \ \ \ $\ell\, \overline{\ell}$ \ \ \
 & \ $\nu_L^{} \overline{\nu_L^{}}$ \
 & \ $\nu_R^{} \overline{\nu_R^{}}$ \
 & \ $\psi_1 \overline{\psi_1}$ \
 & \ $\psi_2 \overline{\psi_2}$ \
 & \ $\psi_3 \overline{\psi_3}$ \
 & \ $\psi_4 \overline{\psi_4}$ \
 & \ ${\mathcal H}^0_1 {\mathcal H}^{0\ast}_1$ \
 & \ ${\mathcal H}^0_2 {\mathcal H}^{0\ast}_2$ \
 & \ $\eta^+ \eta^-$ \
\\\hline
 $0.21$
 & $0.32$
 & $0.16$
 & $0.0059$
 & $0.046$
 & $0.045$
 & $0.044$
 & $0.043$
 & $0.041$
 & $0.038$
 & $0.039$
\end{tabular}
\caption{
 Branching ratios of $Z^\prime$ decays.
}
\label{tab:Zp-decay}
\end{center}
\end{table}
%%<<<<< tab: Z' decay -----

 The light CP-even neutral scalar $h^0$
is made from an $\SU(2)_L$-doublet field $\Phi$
because we take $\cos\theta_0 = 0$.
 The mass $m_{h^0}^{} = 125\,\GeV$ at the benchmark point
is consistent with
$m_{h_\text{SM}^0}^{}
= 125.5 \pm 0.2\,\text{(stat.)}^{+0.5}_{-0.6}\text{(sys.)}\,\GeV$
in the ATLAS experiment~\cite{ATLAS:2013mma}
and
$m_{h^0_\text{SM}}^{}
= 125.7 \pm 0.3\,\text{(stat.)} \pm 0.3\text{(sys.)}\,\GeV$
in the CMS experiment~\cite{CMS:yva}.
 The branching ratio of the invisible decay
$h^0 \to {\mathcal H}^0_1 {\mathcal H}^{0\ast}_1$
is about $7\times 10^{-4}$
for $\lambda_{{\mathcal H}^0_1 {\mathcal H}^0_1 h^0}^{} = 0.001$,
where the recommended value $4.07\,\MeV$~\cite{recommended}
for the total width of $h^0_\SM$ is used.

 For the $Z^\prime$ boson,
the LEP-II bound $m_{Z^\prime}^{}/g_\BL^{} \gtrsim 7\,\TeV$~\cite{LEPZp}
is satisfied at the benchmark point
because of $m_{Z^\prime}^{}/g_\BL^{} = 40\,\TeV$
which we take for a sufficient suppression of $\sigma_\SI$
for the direct search of the dark matter.
 The production cross section of $Z^\prime$
with $g_\BL^{} = 0.1$ and $m_{Z^\prime}^{} = 4000\,\GeV$
is about $0.3\,\text{fb}$ at the LHC with $\sqrt{s}=14\,\TeV$%
~\cite{Ref:BL-Pheno}.%
\footnote{
 The production cross section becomes about $6\,\text{fb}$
if we take $g_\BL^{} = 0.05$ and $m_{Z^\prime}^{} = 2000\,\GeV$.
Notice that
the current bound $m_{Z^\prime}^{} \gtrsim 3\,\TeV$ at the LHC~\cite{LHCZp}
is for the case where
the gauge coupling for $Z^\prime$ is
the same as the one for $Z$,
namely $g_\BL^{} \simeq 0.7$.
}
 Decay branching ratios of $Z^\prime$ are shown
at the benchmark point in Table~\ref{tab:Zp-decay}.

%%---- tab: nuR decay >>>>>
\begin{table}[t]
\begin{center}
\begin{tabular}{c|c|c|c}
 \ \ \ $W^+ \ell^- + W^- \ell^+$ \ \ \
 & \ \ \ $Z \nu_L^{} + Z \overline{\nu_L^{}}$ \ \ \
 & \ $h^0 \nu_L^{} + h^0 \overline{\nu_L^{}}$ \
 & \ $H^0 \nu_L^{} + H^0 \overline{\nu_L^{}}$ \
\\\hline
 $0.56$
 & $0.28$
 & $0.16$
 & $0$
\end{tabular}
\caption{
 Branching ratios of $\nu_R^{}$ decays.
}
\label{tab:nuR-decay}
\end{center}
\end{table}
%%<<<<< tab: nuR decay -----

 Decays of $\psi_i$ are dominated by
$\psi_i \to \nu_R^{} {\mathcal H}^0_1$
with the Yukawa coupling constants $h_{i1}$
because $y_{\ell i}^{}$ for $\psi_i \to \ell^\pm \eta^\mp$
are small in order to satisfy the $\mu \to e\gamma$ constraint.
 The ${\mathcal H}^0_2$~($\simeq \eta^0$)
decays into $h^0 {\mathcal H}^0_1$ via
the trilinear coupling constant $\mu_3^{}$.
 The main decay mode of the charged scalar is
$\eta^\pm \to W^\pm {\mathcal H}^0_1$
through the mixing $\theta_0^\prime$
between $\eta^0$ and $s^0$.

 In this model,
$\nu_R^{}$ is not the dark matter
and can decay into the SM particles.
 Decay branching ratios for $\nu_R^{}$
are shown in Table~\ref{tab:nuR-decay}.
 The decay into $H^0$ is forbidden
because it is heavier than $\nu_R^{}$
at the benchmark point.
 Since the $\BL$ charge of $\nu_R^{}$ is rather small,
$\nu_R^{}$ is not produced directly from $Z^\prime$.
 However,
$\nu_R^{}$ can be produced
through the decays of $\psi_i$.
 As a result,
about $18\,\%$ of $Z^\prime$ produces $\nu_R^{}$.
For $\nu_R^{} \to W \ell$~($56\,\%$)
followed by the hadronic decay of $W$~($68\,\%$),
the $\nu_R^{}$ would be reconstructed.
 In this model,
an invariant mass of
a pair of the reconstructed $\nu_R^{}$
is not at $m_{Z^\prime}^{}$
in contrast with a naive model
where only three $\nu_R^{}$ with $\BL = -1$
are introduced to the SM.%
\footnote{
In the naive model
with $m_{Ra}^{} = 250\,\GeV$~(degenerate) and $m_{Z^\prime}^{} = 4\,\TeV$,
the decay branching ratios of $Z^\prime$
into
$\{ q\overline{q},\, \ell\overline{\ell},\,
\nu_L^{}\overline{\nu_L^{}},\, \nu_R^{} \overline{\nu_R^{}} \}$
are $\{ 0.25,\, 0.38,\, 0.19,\, 0.19 \}$.
}
 This feature of $\nu_R$
also enables us to distinguish this model
from the previous model in Ref.~\cite{Ref:KNS}
where $\nu_R$ with $\BL = 1$ can be directly produced
by the $Z^\prime$ decay.

%%%%%%%%%%%%%%%%%%%%%%%%%%%
%%%  sec: conclusion   %%%%
%%%%%%%%%%%%%%%%%%%%%%%%%%%
\section{Conclusions}
\label{sec:concl}

 We have proposed the model
which is an improved version of the model in Ref.~\cite{Ref:KNS}
by considering anomaly cancellation
of the $\U(1)_\BL$ gauge symmetry.
 We have shown that
there are four anomaly-free cases
of $\BL$ charge assignment,
and three of them have an unbroken global $\U(1)_\DM$ symmetry
(one of the three is not acceptable
because two neutrinos become massless).
 The $\U(1)_\DM$ guarantees
that the lightest $\U(1)_\DM$-charged particle is stable
such that it can be regarded as a dark matter candidate.
 The spontaneous breaking of the $\U(1)_\BL$ symmetry
generates new fermion mass terms
which do not exist in the SM;
namely, the Dirac mass term of neutrinos,
the Majorana mass term of $\nu_R^{}$,
and masses of new fermions $\psi$.
 Especially,
the Dirac mass term of neutrinos
is generated at the one-loop level
where the dark matter candidate involved in the loop.
 Tiny neutrino masses are obtained
at the two-loop level.
 The case of the fermion dark matter is excluded,
and the lightest $\U(1)_\DM$-charged scalar ${\mathcal H}_1^0$
should be the dark matter in this model.
 We have found a benchmark point of model parameters
which satisfies current constraints from
neutrino oscillation data, lepton flavor violation searches,
the relic abundance of the dark matter,
direct searches for the dark matter,
and the LHC experiments.

 By virtue of the radiative mechanism
for the Dirac mass term of neutrinos,
very heavy $\nu_R^{}$ are not required for tiny neutrino masses.
 Therefore,
$\nu_R^{}$ would be produced at the LHC\@.
 In contrast to a naive model
where three $\nu_R^{}$ have $\BL = -1$
and the model in Ref.~\cite{Ref:KNS} where
$\nu_R^{}$ have $\BL = 1$,
the $\nu_R^{}$ with $\BL = -1/3$ in this model
cannot be directly produced by the $Z^\prime$ decay,
but can be produced by the cascade decay
$Z^\prime \to \psi_i \overline{\psi}_i
\to \nu_R^{} \overline{\nu_R^{}} {\mathcal H}_1^0 {\mathcal H}_1^{0\ast}$.
 The invariant mass distribution of $\nu_R^{} \overline{\nu_R^{}}$
does not take a peak at $m_{Z^\prime}^{}$,
which could be a characteristic signal
of this kind of models with the unusual $\BL$ charge of $\nu_R^{}$.

%%%%%%%%%%%%%%%%%%%%%%%%%%%%%%%%%%%%%%%%%%%
%%%%%%%%%  acknowledgments  %%%%%%%%%%%%%%%
%%%%%%%%%%%%%%%%%%%%%%%%%%%%%%%%%%%%%%%%%%%
\begin{acknowledgments}
 We would like to thank Takehiro Nabeshima for fruitful discussions.
 T.M.\ also thanks Naoki Machida for useful discussions.
 The work of S.K.\ was supported, in part,
by Grant-in-Aid for Scientific research
from the Japan Society for the Promotion of Science~(JSPS)
Nos.~22244031 and 24340046,
and from the Ministry of Education,
Culture, Sports, Science and Technology~(MEXT), Japan,
No.~23104006.
\end{acknowledgments}

%%%%%%%%%%  appendix  %%%%%%%%%%%%%%%
\appendix
\section{Loop Integration}
\label{sec:loop}

 A loop function $(I_1)_{ija}$ in eq.~\eqref{eq:mnu}
can be expressed as
%%------------------------->>>>>
\begin{eqnarray}
(I_1)_{ija}
&\equiv&
 -
 \frac{ ( 8\pi^2 \sin{2\theta_0^\prime} )^2 m_{\psi_i}^{} m_{\psi_j}^{} }
      { (m_R^{})_a^2 }
 \left[
  \int\!\!\frac{d^4p}{(2\pi)^4}
  \frac{1}{ p^2 - m_{\psi_i}^2 }
  \left\{
   \frac{1}{ p^2 - m_{{\mathcal H}^0_1}^2 }
   -
   \frac{1}{ p^2 - m_{{\mathcal H}^0_2}^2 }
  \right\}
 \right]
\nonumber\\
&&\hspace*{40mm}
 \times
 \left[
  \int\frac{d^4q}{(2\pi)^4}
  \frac{1}{ q^2 - m_{\psi_j}^2 }
  \left\{
   \frac{1}{ q^2 - m_{{\mathcal H}^0_1}^2 }
   -
   \frac{1}{ q^2 - m_{{\mathcal H}^0_2}^2 }
  \right\}
 \right]
\nonumber\\
&=&
 \frac{
       m_{\psi_i}^{} m_{\psi_j}^{}
       ( m_{{\mathcal H}^0_1}^2 - m_{{\mathcal H}^0_2}^2 )^2
       \sin^2{2\theta_0^\prime}
      }
      { 4 (m_R^{})_a^2 }
 \Bigl\{
  C_0( 0, 0, m_{\psi_i}^{}, m_{{\mathcal H}^0_1}^2, m_{{\mathcal H}^0_2}^2 )
\nonumber\\
&&\hspace*{60mm}
 \times
  C_0( 0, 0, m_{\psi_j}^{}, m_{{\mathcal H}^0_1}^2, m_{{\mathcal H}^0_2}^2 )
 \Bigr\} ,
\end{eqnarray}
%%<<<<<-------------------------
where the $C_0$ function~\cite{Passarino:1978jh} is given by
%%------------------------->>>>>
\begin{eqnarray}
&&\hspace*{-10mm}
C_0( 0, 0, m_0^2, m_1^2, m_2^2 )
\nonumber\\
&&\hspace*{-2mm}
\equiv
 \frac{1}{ ( m_0^2 - m_1^2 ) ( m_1^2 - m_2^2 ) ( m_2^2 - m_0^2 ) }
 \Bigl\{
  m_0^2 m_1^2 \ln\frac{m_0^2}{m_1^2}
  + m_1^2 m_2^2 \ln\frac{m_1^2}{m_2^2}
  + m_2^2 m_0^2 \ln\frac{m_2^2}{m_0^2}
 \Bigr\} .
\end{eqnarray}
%%<<<<<-------------------------

 On the other hand,
another loop function $(I_2)_{ija}$ in eq.~\eqref{eq:mnu}
is given by
%%------------------------->>>>>
\begin{eqnarray}
(I_2)_{ija}
&\equiv&
 ( 8\pi^2 \sin{2\theta_0^\prime} )^2 m_{\psi_i}^{} m_{\psi_j}^{}
 \nonumber\\
&&\times
 \int\!\!\!\int\frac{d^4p}{(2\pi)^4}\frac{d^4q}{(2\pi)^4}
 \left\{
  \frac{1}{ p^2 - m_{{\mathcal H}^0_1}^2 }
  -
  \frac{1}{ p^2 - m_{{\mathcal H}^0_2}^2 }
 \right\}
 \frac{1}{ p^2 - m_{\psi_i}^2 }
 \nonumber\\
&&\times
 \frac{1}{ (p+q)^2 - (m_R^{})_a^2 }
 \left\{
  \frac{1}{ q^2 - m_{{\mathcal H}^0_1}^2 }
  -
  \frac{1}{ q^2 - m_{{\mathcal H}^0_2}^2 }
 \right\}
 \frac{1}{ q^2 - m_{\psi_j}^2 }
\nonumber\\
&=&
 ( 8\pi^2 \sin{2\theta_0^\prime} )^2 m_{\psi_i}^{} m_{\psi_j}^{}
 \nonumber\\
&&\times
 \Bigl[
  I( m_{{\mathcal H}^0_1}^{}, m_{\psi_i}^{} |
     m_{{\mathcal H}^0_1}^{}, m_{\psi_j}^{} | (m_R^{})_a )
  - I( m_{{\mathcal H}^0_1}^{}, m_{\psi_i}^{} |
       m_{{\mathcal H}^0_2}^{}, m_{\psi_j}^{} | (m_R^{})_a )
 \nonumber\\
&&\hspace*{10mm}{}
  - I( m_{{\mathcal H}^0_2}^{}, m_{\psi_i}^{} |
       m_{{\mathcal H}^0_1}^{}, m_{\psi_j}^{} | (m_R^{})_a )
  + I( m_{{\mathcal H}^0_2}^{}, m_{\psi_i}^{} |
       m_{{\mathcal H}^0_2}^{}, m_{\psi_j}^{} | (m_R^{})_a )
 \Bigr] ,
\end{eqnarray}
%%<<<<<-------------------------
where
%%------------------------->>>>>
\begin{eqnarray}
&&
I( m_{11}^{}, m_{12}^{}, \cdots, m_{1n_1^{}}^{} |
   m_{21}^{}, m_{22}^{}, \cdots, m_{2n_2^{}}^{} |
   m_{31}^{}, m_{32}^{}, \cdots, m_{3n_3^{}}^{} )
\nonumber\\
&&\hspace*{10mm}
\equiv
 \int\!\!\frac{d^4p_E^{}}{(2\pi)^4}
 \int\!\!\frac{d^4q_E^{}}{(2\pi)^4}\,
 \prod_{i=1}^{n_1^{}}
 \prod_{j=1}^{n_2^{}}
 \prod_{k=1}^{n_3^{}}
 \frac{1}{ p_E^2 + m_{1i}^2 }\,
 \frac{1}{ q_E^2 + m_{2j}^2 }\,
 \frac{1}{ (p_E^{}+q_E^{})^2 + m_{3k}^2 } .
\end{eqnarray}
%%<<<<<-------------------------
 We can use the following results~\cite{Ref:2loop}:
%%------------------------->>>>>
\begin{eqnarray}
&&
I(m_{11}, m_{12}| m_{21}, m_{22}| m_3)
\nonumber\\
&&\hspace*{10mm}
=
 \frac{
        I( m_{12}^{} | m_{22}^{} | m_3^{} )
        - I( m_{11}^{} | m_{22}^{} | m_3^{} )
        - I( m_{12}^{} | m_{21}^{} | m_3^{} )
        + I( m_{11}^{} | m_{21}^{} | m_3^{} )
      }
      {
        (16\pi^2)^2
        ( m_{11}^2 - m_{12}^2 )
        ( m_{21}^2 - m_{22}^2 )
      } ,
\end{eqnarray}
%
%-----------------------
%
\begin{eqnarray}
I( m_1 | m_2 | m_3 )
&=&
 - m_1^2\, f\!\left( \frac{m_2^2}{m_1^2}, \frac{m_3^2}{m_1^2} \right)
 - m_2^2\, f\!\left( \frac{m_1^2}{m_2^2}, \frac{m_3^2}{m_2^2} \right)
 - m_3^2\, f\!\left( \frac{m_1^2}{m_3^2}, \frac{m_2^2}{m_3^2} \right) ,
\end{eqnarray}
where
\begin{eqnarray}
f(x, y)
&\equiv&
 - \frac{1}{\,2\,} (\ln x) (\ln y)
 - \frac{1}{\,2\,} \left( \frac{ x+y-1 }{D} \right)
\nonumber\\
&&
\times
 \Bigl\{
  \Li_2\!\left( \frac{-x_-}{y_+} \right)
  + \Li_2\!\left( \frac{-y_-}{x_+} \right)
  - \Li_2\!\left( \frac{-x_+}{y_-} \right)
  - \Li_2\!\left( \frac{-y_+}{x_-} \right)
\nonumber\\
&&\hspace*{10mm}
{}+ \Li_2\!\left( \frac{y-x}{x_-} \right)
  + \Li_2\!\left( \frac{x-y}{y_-} \right)
  - \Li_2\!\left( \frac{y-x}{x_+} \right)
  - \Li_2\!\left( \frac{x-y}{y_+} \right)
 \Bigr\} ,
\end{eqnarray}
and
\begin{eqnarray}
D
&\equiv&
 \sqrt{ 1 - 2 (x+y) + (x-y)^2 } ,
\\
%
%---------------------
%
x_\pm
&\equiv&
 \frac{1}{\,2\,}
 \left( 1 - x + y \pm D \right) , \qquad
%
%---------------------
%
y_\pm
\equiv
 \frac{1}{\,2\,}
 \left( 1 + x - y \pm D \right) ,
\end{eqnarray}
and the dilog function $\Li_2(x)$ is defined as
\begin{eqnarray}
\Li_2(x)
\equiv
 -\int_0^x\!\!dt\, \frac{ \ln(1-t) }{t} .
\end{eqnarray}
%%<<<<<-------------------------

\section{Ansatz for benchmark point}
\label{sec:ansatz}

 The symmetric matrix $A_{ij}$ in eq.~\eqref{eq:loop-matrix}
can be diagonalized
by an orthogonal matrix $X$ as
%%----- eq: Diagonalize A >>>>>
\begin{eqnarray}
X A X^T = \text{diag}(a_1, a_2, a_3, a_4).
\end{eqnarray}
%%<<<<< eq: Diagonalize A-----
 It is clear that
a Yukawa matrix $f_{\ell i}$
of the following structure
satisfies constraints from neutrino oscillation data:
%%----- eq: Ansatz >>>>>
\begin{eqnarray}
f
&=&
 16\pi^2\,
 U_\MNS
 \begin{pmatrix}
  \sqrt{ \frac{ m_1 }{ |a_1| } }
   & 0
   & 0
   & 0\\
  0
   & \sqrt{ \frac{ m_2 }{ |a_2| } }
   & 0
   & 0\\
  0
   & 0
   & \sqrt{ \frac{ m_3 }{ |a_3| } }
   & 0
 \end{pmatrix}
 X ,
\end{eqnarray}
%%<<<<< eq: Ansatz -----
where Majorana phases are given by
$\alpha_i = \text{arg}(a_i)$.
 We used
%%----- eq: Matrix X >>>>>
\begin{eqnarray}
X
=
 \begin{pmatrix}
   0.520 & -0.520 & -0.474 &  0.484\\
  -0.712 & -0.284 &  0.165 &  0.621\\
  -0.425 & -0.476 & -0.522 & -0.566\\
   0.206 & -0.650 &  0.689 & -0.244
 \end{pmatrix} ,
%X
%=
% \begin{pmatrix}
%   0.520145 & -0.519843 & -0.474162 &  0.484131\\
%  -0.711653 & -0.283881 &  0.164763 &  0.62114\\
%  -0.425038 & -0.475682 & -0.52244  & -0.565796\\
%   0.205771 & -0.650308 &  0.689261 & -0.244289
% \end{pmatrix} ,
\end{eqnarray}
%%<<<<< eq: Matrix X -----
where $0< a_4 < a_1 < a_2 < a_3$.
 The ordering of eigenvalues $a_i$
is preferred to suppress $y_{\ell i}^{}$
(in order to satisfy a constraint from $\mu\to e\gamma$ search)
for the normal mass ordering for neutrinos~($m_1 < m_2 < m_3$).
 With this ansatz,
small neutrino masses are preferred
to suppress $\BR(\mu\to e\gamma)$.

\end{document}